\UseRawInputEncoding
\documentclass[amsmath,amssymb,prb,twocolumn,superscriptaddress]{revtex4-2}
\usepackage[compatibility=false]{caption}
\usepackage{subcaption}
\usepackage[export]{adjustbox}
\usepackage[normalem]{ulem}
\usepackage[table,dvipsnames]{xcolor}

\usepackage[version=3]{mhchem} 
\usepackage{comment}
\usepackage{enumitem}
\usepackage{amsmath}
\usepackage{supertabular} 

\usepackage{booktabs}  
\usepackage{array}     
\usepackage{ragged2e}   
\usepackage{tabularx}   
\usepackage{setspace, colortbl}
\definecolor{tblgray}{gray}{0.85}

\newcolumntype{S}{>{\RaggedRight\arraybackslash}m{0.40\textwidth}}
\newcolumntype{E}{>{\RaggedRight\arraybackslash}m{1\textwidth}}
\newcolumntype{K}{>{\RaggedRight\arraybackslash}m{1.1\textwidth}}

\definecolor{lightyellow}{rgb}{1,1,0.8}
\definecolor{darkgray}{rgb}{0.7,0.7,0.7}
\definecolor{lightgray}{rgb}{0.9,0.9,0.9}

\definecolor{MXenesDark}{rgb}{0.898, 0.620, 0.867}
\definecolor{MXenesLight}{rgb}{0.949, 0.812, 0.933}
\definecolor{XenesDark}{rgb}{0.325, 0.620, 0.73}
\definecolor{ChalDark}{rgb}{0.557, 0.851, 0.451}
\definecolor{ChalLight}{rgb}{0.839, 0.945, 0.800}
\definecolor{HalidesDark}{rgb}{1.000, 0.808, 0.219}
\definecolor{OxidesDark}{rgb}{0.925, 0.369, 0.451}
\definecolor{OxidesLight}{rgb}{0.973, 0.773, 0.804}
\definecolor{MixedDark}{rgb}{0.949, 0.667, 0.518}

\newcolumntype{L}{>{\RaggedRight\arraybackslash}m{0.38\textwidth}}
\newcolumntype{R}{>{\RaggedRight\arraybackslash}m{0.57\textwidth}}

\begin{document}

\begin{abstract}
The past decade has seen rapid growth in the number of experimentally realized two-dimensional (2D) materials with diverse chemical and physical properties. However, information on their crystal structure, synthesis routes, and measured or predicted properties, remains scattered across thousands of publications. Here we consolidate this fragmented knowledge by establishing X2DB -- an open infrastructure that integrates experimental and computational data on 2D materials. Using extensive literature mining and direct community uploads, we identify 370 unique 2D materials that have been realized in monolayer or few-layer form, and link them to their digital counterparts in computational databases, enabling consistent \emph{ab initio} characterization of their properties across monolayer, bilayer and bulk forms. We describe the structure and content of the database highlighting its support for community uploads, illustrate how it can be used to generate new scientific insight and introduce a hierarchical classification of the known set of 2D materials. Our work provides a foundation for the integration and cross-fertilization of experimental and theoretical knowledge, opening new avenues for data-driven, predictive synthesis of novel 2D materials. 
\end{abstract}

\title{Large-scale Integration of Experimental and Computational Data for 2D Materials}

\author{Mohammad A. Akhound}
\affiliation{CAMD, Department of Physics, Technical University of Denmark, DK - 2800 Kongens Lyngby, Denmark}
\email{akhound@dtu.dk, thygesen@fysik.dtu.dk}

\author{Tara M. Boland}
\affiliation{CAMD, Department of Physics, Technical University of Denmark, DK - 2800 Kongens Lyngby, Denmark}

\author{Mikkel O. Sauer}
\affiliation{CAMD, Department of Physics, Technical University of Denmark, DK - 2800 Kongens Lyngby, Denmark}

\author{Matthias Batzill}
\affiliation{Department of Physics, University of South Florida, Tampa, FL 33620, USA.}

\author{Moses A. Bokinala}
\affiliation{Materials Science and Engineering Department and A. J. Drexel Nanomaterials Institute, Drexel University, 3141 Chestnut Street, Philadelphia, Pennsylvania 19104, United States}

\author{Stela Canulescu}
\affiliation{Department of Electrical and Photonics Engineering, Technical University of Denmark, Roskilde 4000, Denmark}

\author{Yury Gogotsi}
\affiliation{Materials Science and Engineering Department and A. J. Drexel Nanomaterials Institute, Drexel University, 3141 Chestnut Street, Philadelphia, Pennsylvania 19104, United States}

\author{Philip Hofmann}
\affiliation{Department of Physics and Astronomy, Aarhus University, 8000 Aarhus C, Denmark}

\author{Andras Kis}
\affiliation{Institute of Electrical and Microengineering, École Polytechnique Fédérale de Lausanne (EPFL), Lausanne, Switzerland}

\author{Jiong Lu}
\affiliation{Institute for Functional Intelligent Materials,
National University of Singapore}

\author{Thomas Michely}
\affiliation{Physikalisches Institut, Universität zu
Köln, D-50937 Köln, Germany}

\author{S\o ren Raza}
\affiliation{Department of Physics, Technical University of Denmark, DK - 2800 Kongens Lyngby, Denmark}

\author{Wencai Ren}
\affiliation{Shenyang National Laboratory for Materials Science, Institute of Metal Research, Chinese Academy of Sciences, 72
Wenhua Road, Shenyang 110016, P. R. China.}

\author{Joshua A. Robinson}
\affiliation{Materials Science and Engineering at the Pennsylvania State University, United States}

\author{Zdenek Sofer}
\affiliation{Department of Inorganic Chemistry, University of Chemistry and Technology Prague, Technická 5, 166 28 Prague 6, Czech Republic}

\author{Jing H. Teng}
\affiliation{Institute of Materials Research and Engineering (IMRE), Agency for Science, Technology and Research (A*STAR), Singapore, 138634 Singapore}

\author{S\o ren Ulstrup}
\affiliation{Department of Physics and Astronomy, Aarhus University, 8000 Aarhus C, Denmark}

\author{Meng Zhao}
\affiliation{Institute of Materials Research and Engineering (IMRE), Agency for Science, Technology and Research (A*STAR), Singapore, 138634 Singapore}

\author{Xiaoxu Zhao}
\affiliation{School of Materials Science and Engineering, Peking University, Beijing 100871, China}

\author{Jens J. Mortensen}
\affiliation{CAMD, Department of Physics, Technical University of Denmark, DK - 2800 Kongens Lyngby, Denmark}

\author{Thomas Olsen}
\affiliation{CAMD, Department of Physics, Technical University of Denmark, DK - 2800 Kongens Lyngby, Denmark}

\author{Kristian S. Thygesen}
\affiliation{CAMD, Department of Physics, Technical University of Denmark, DK - 2800 Kongens Lyngby, Denmark}
\email{thygesen@fysik.dtu.dk}

\maketitle

\section{Introduction}
Two-dimensional (2D) materials have emerged as a distinct class of materials with exceptional physical properties and broad technological potential\cite{ren20252d} that spans from catalysis \cite{He2024AdvancesConversion} and energy storage\cite{Anasori20172DStorage} to photonics \cite{doi:10.1021/acsphotonics.5c00353}, electronics \cite{Song2018Two-DimensionalApplications}, and quantum devices\cite{liu20192d}. The reduced dimensionality simplifies the space of possible 2D materials compared to bulk solids -- for example, there are only 80 layer groups versus 231 three-dimensional space groups\cite{fu2024symmetry}. Yet, the known 2D materials exhibit vast diversity in electronic behavior covering band insulators\cite{Song2010LargeLayers}, Mott insulators\cite{ma2016metallic}, semiconductors\cite{Wang2012ElectronicsDichalcogenides}, semi-metals\cite{Novoselov2004ElectricFilms},  metals\cite{zhao20212d}, superconductors\cite{wang2017high}, ferroelectrics\cite{wang2023towards}, and various types of topological\cite{xu2018electrically} and magnetic phases\cite{gibertini2019magnetic}.

The high pace at which new 2D materials and related physical phenomena have been discovered, has made it difficult to organize the resulting knowledge efficiently. In contrast to bulk materials, where more than a century of research has been consolidated in handbooks and digital databases like the Inorganic Crystal Structure Database (ICSD)\cite{zagorac2019recent} and the Materials Project\cite{jain2013commentary}, experimental data on 2D materials remain scattered across thousands of publications. While several computational databases on 2D materials exist\cite{Haastrup2018TheCrystals,campi2023expansion,zhou20192dmatpedia}, no comparable resource exist for experimental realizations. 

This lack of structured experimental data creates a critical gap: of the thousands of 2D materials predicted by  computations\cite{Haastrup2018TheCrystals,mounet2018two}, only a small subset has been synthesized. This discrepancy arises from several factors. First, the synthesis process of 2D materials can be complex and require highly specific conditions\cite{Wang2024SynthesisSubstitution} - both for bottom-up growth and for top-down exfoliation\cite{Zhao2024ElectrochemicalGraphene}. In addition, the experimental techniques used to synthesize and characterize 2D materials are still evolving and often involve advanced equipment that is not universally accessible\cite{Choi2022Large-scaleIndustrialization}. Computational predictions, while valuable, are often based on simple descriptors that are easy to calculate but rely on assumptions that may not fully represent the real-world situation\cite{Ryu2022UnderstandingLearning}. Most notably, the decomposition energy (aka the energy above the convex hull) is widely used as a descriptor for thermodynamic stability even thought it neglect effects of temperature/entropy, substrates interactions, and chemical environments. Despite the common use of this and other simple descriptors (e.g. phonons for dynamical stability\cite{manti2023exploring} and interlayer binding energy for exfoliability\cite{mounet2018two}), their importance for stability under realistic conditions and relation to synthesizability, remains unclear. We note in passing that more advanced stability descriptors have been proposed to predict e.g. chemical exfoliability\cite{bjork2024two} and environmental stability of 2D materials\cite{americo2025predicting}.   
   
The gap between theoretically predicted materials and experimental realizations raises important questions regarding the relevance and predictive power of established computational materials-discovery approaches and underscores the challenges of synthesizing novel 2D materials. \cite{Lin2023RecentApplications, Zhang20222DElectro/Photocatalysis} An essential first step towards answering these questions and overcoming the challenges, is to establish a systematic overview of the current landscape of experimentally realized 2D materials and connect them to relevant computations.

Earlier work in this direction has focused on known bulk compounds with potential for exfoliation. Following initial work by Leb{\`e}gue \emph{et al.}\cite{lebegue2013two}, several groups employed topological screening of experimental crystal structure databases in combination with DFT calculations, concluding that nearly 2000 known bulk phases can potentially be mechanically exfoliated.\cite{cheon2017data,ashton2017topology,mounet2018two} More recently,
Björk \emph{et al.} used the formation of MXenes from MAX phases as a guiding example to computationally identify non-vdW layered compounds potentially exfoliable by selective etching.\cite{bjork2024two} However, aside from a few well known reference cases, these studies did not assess which materials have actually been realized in atomically thin form. 

\begin{figure*}[t!]
 \centering
 \includegraphics[height=8.5cm]{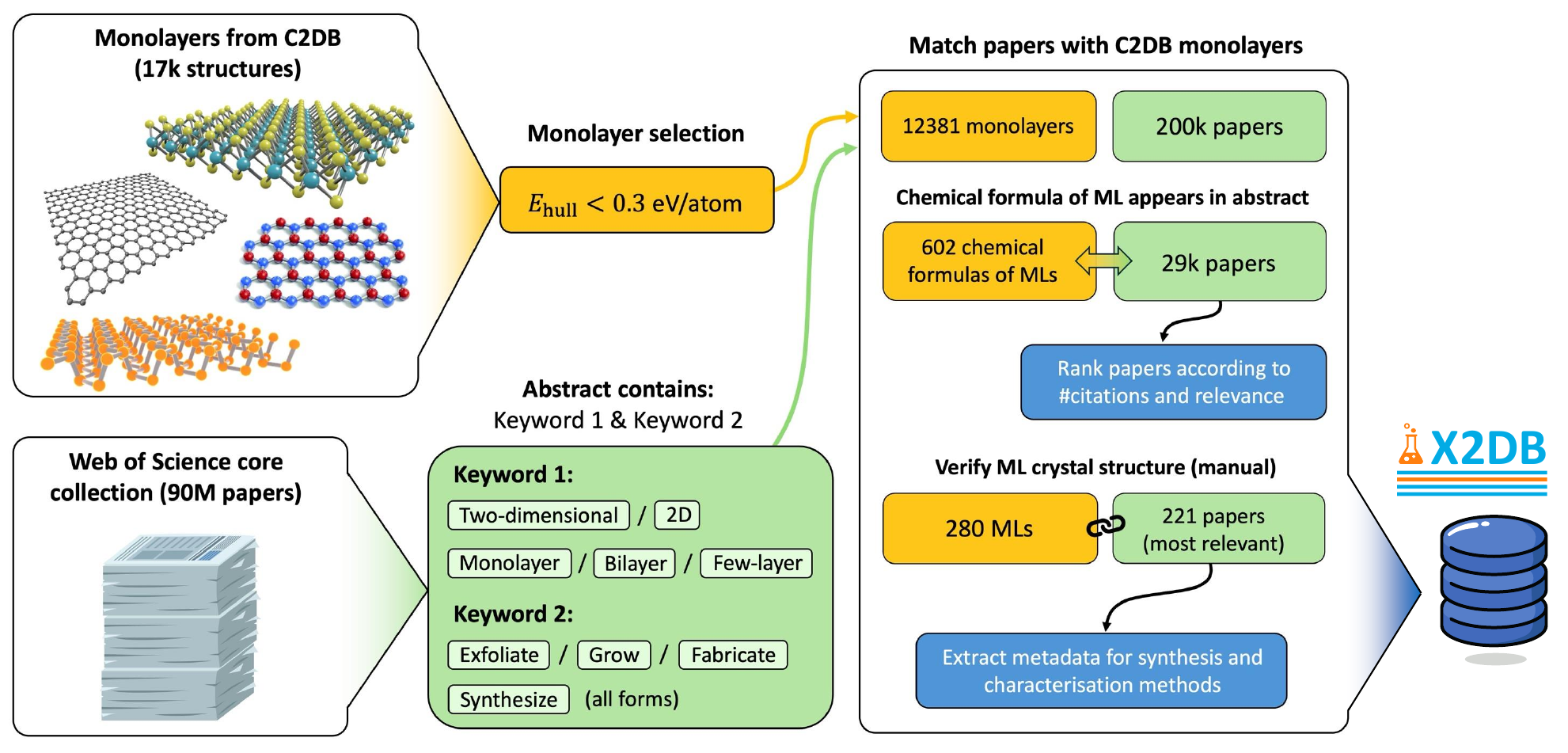}
 \caption{\textbf{Workflow used to identify experimentally reported 2D materials and link them to C2DB.} Relevant publications are identified by searching the abstract for specific keywords that indicate that the paper reports on the synthesis or exfoliation of a material in atomically thin form. The materials reported in the papers are linked to structures in the C2DB monolayer database using manual inspection of the crystal structure. Finally, the information describing the reported material and the employed synthesis and characterization methods, is extracted and classified according to the 2D materials taxonomy.}
 \label{fig:workflow}
\end{figure*}

In this work, we provide the first systematic mapping of experimentally realized 2D materials and furthermore link them directly to their counterparts in open computational databases. By screening the 2D materials literature using a combination of filtering, ranking, and manual structure verification, we identify 270 unique materials that have been synthesized in few-layer form and whose chemical formula appears in the Computational 2D Materials Database (C2DB). To enable consistent descriptions, we introduce a unified taxonomy for 2D materials covering crystal structure, sample morphology, synthesis route, and characterization methods, and we curate this information in the open-access Experimental 2D Materials Database (X2DB). Via its full integration with computational monolayer, bilayer, and bulk databases, the X2DB allows seamless exploration of consistently calculated properties across length scales (monolayer, bilayer, bulk) for experimentally relevant materials. Importantly, X2DB also supports community uploads, allowing that the database continues to expand and remains a living resource for the field. Indeed, community submissions have already expanded the X2DB catalogue by approximately 100 additional unique 2D materials.

\begin{figure}
 \includegraphics[height=5cm]{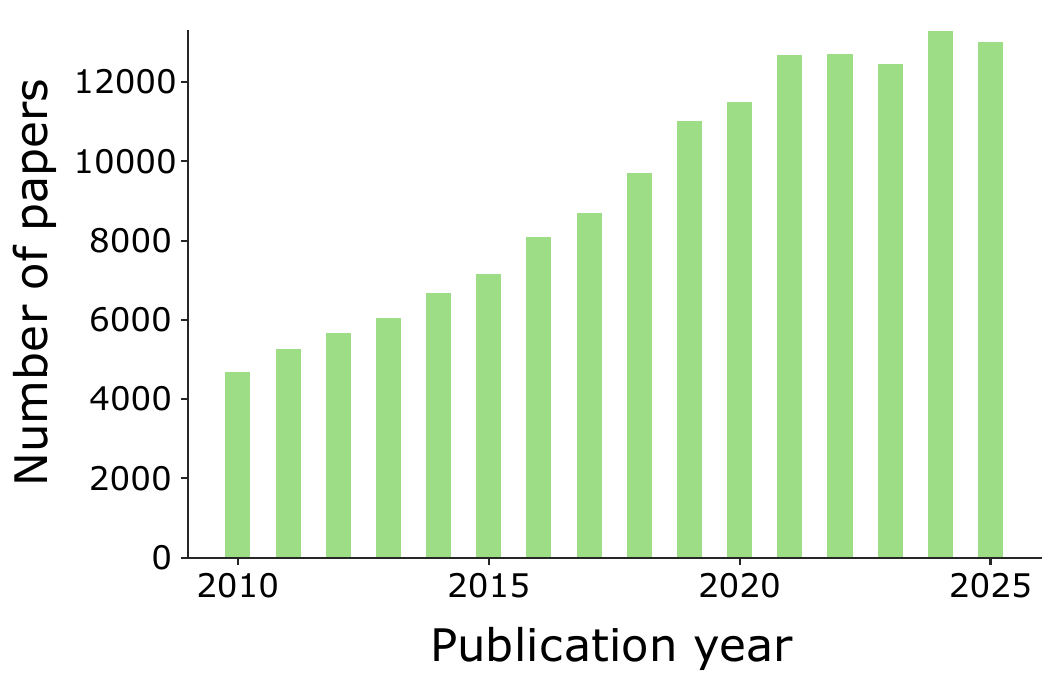}
 \caption{Papers published on experimentally realized 2D materials since 2010. More precisely, the graph shows the 200k publications selected by the search criteria in the green box in Fig. \ref{fig:workflow}, sorted according to publication year.}
 \label{fig:papers_per_year}
\end{figure}

\section{Experimentally realized 2D materials}
The workflow used to identify experimentally realized 2D materials and connect them to monolayers in the C2DB is illustrated in Fig.\ref{fig:workflow}. The process began by filtering approximately 90 million scientific papers from the Web of Science Core Collection by selecting those whose abstracts contained keywords indicative of experimental studies on 2D materials (see green box in Fig.\ref{fig:workflow}). This initial filtering yielded around 200,000 papers; their distribution over publication years shown in Fig.~\ref{fig:papers_per_year}.
Next, we refined the selection by retaining only papers whose abstracts include a chemical formula matching a stable monolayer in the C2DB (here stable means an energy above the convex hull below 0.3 eV/atom). This filters out reports on materials that are unlikely to be found in the C2DB, and reduced the dataset to approximately 29,000 papers. We note that this filter induces a bias towards theoretically predicted materials with the risk of missing genuinely new (not theoretically explored) materials. On the other hand, it ensures that the experimental reports can be linked to relevant computational data, which is a key feature of our work. 

The 29,000 papers were then ranked according to a set of relevance criteria, including citation count, publication year, and the presence of density functional theory (DFT) calculations supporting the experiments. Starting from the top-ranked papers, we performed manual inspection of each paper to verify that it indeed contains experimental synthesis and characterization of a 2D material in atomically thin form.

Following this manual verification, we identified 280 unique materials reported in 221 papers (some papers report multiple materials). For each material, we performed a detailed comparison of the reported experimental data with the structural polytypes of the relevant chemical formula available in the C2DB to verify the crystal structure. Since the reported experimental data do not always allow an unambiguous identification of the structure, we assign a \emph{confidence level} to the pairing between an experimentally realized material and a C2DB monolayer structure, taking values 1 (low), 2 (medium), or 3 (high). Out of the 280 materials, 163 were assigned a confidence level of 3, indicating that their structural phase could be identified without ambiguity.

The literature mining described above was performed to populate the database with a diverse initial set of materials. As a second step, the database was opened up to accept registrations of materials/publications from a limited group of experimental researchers who also provided feedback on the infrastructure design and user interface. At the time of writing, the X2DB holds 550 entries covering 370 unique chemical formulas of which 210 have been matched with a monolayer in C2DB with highest confidence level.    

On this basis, we can conclude that at least 370 distinct materials have so far been experimentally realized in monolayer or few-layer form. This number should, for obvious reasons be considered as a lower bound. An overview of the materials (the subset identified after the initial literature mining step) is provided in the Supplementary Information (SI), along with references to the corresponding publications and a summary of the experimental methods used. Additionally, each material is linked to its monolayer in the C2DB, where a variety of computed properties are available. We note that this information, and more, is also available via the X2DB website.


\section{The 2D Materials Taxonomy}
\begin{figure*}
 \centering
 \includegraphics[height=7.5cm]{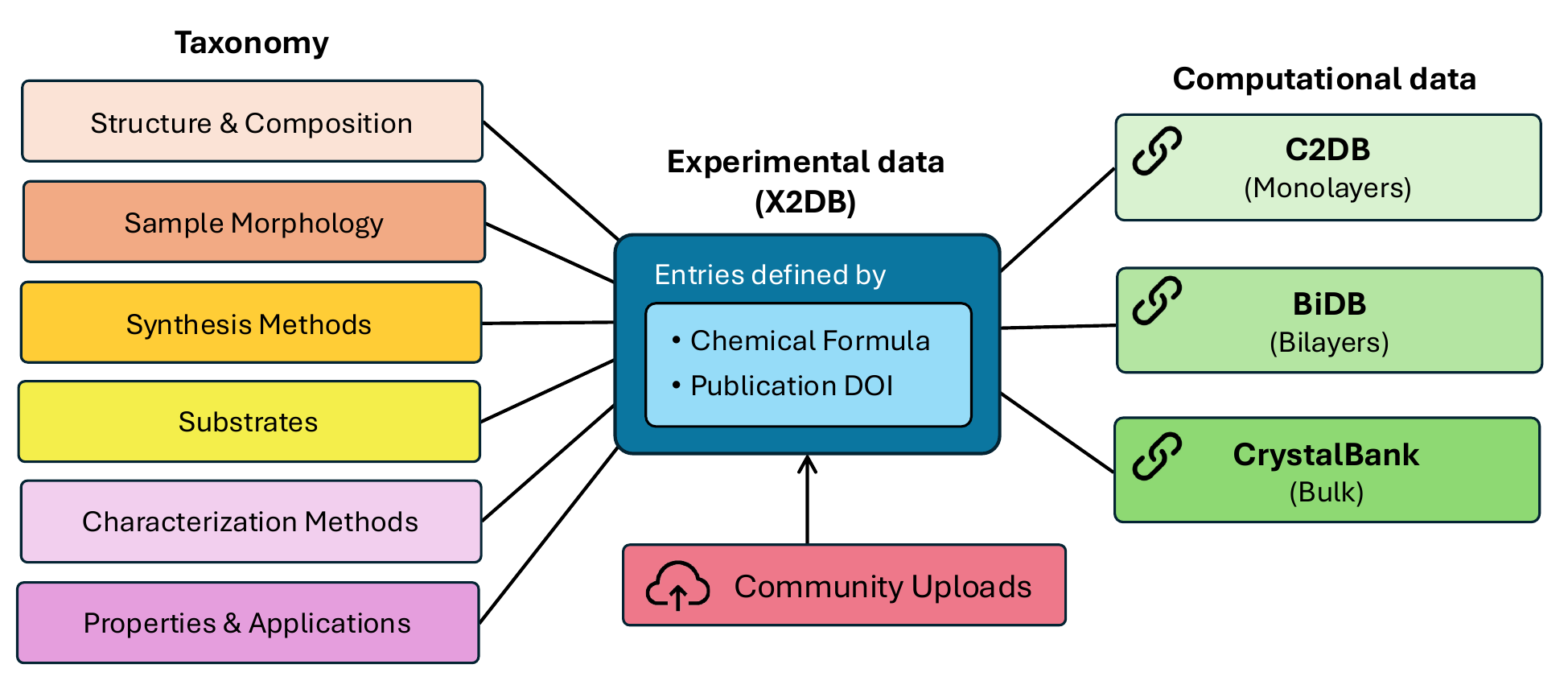}
 \caption{\textbf{The Experimental 2D Materials Database}. The taxonomy (left) provides a controlled language for describing 2D materials. Colored boxes represent the taxonomy's top-level categories each containing sub-categories and predefined labels as detailed in the SI. Experimental materials can be linked to corresponding structures in computational databases of monolayers, bilayers, and bulk solids. An associated confidence level (number between 1 and 3) is used to indicate the reliability of the link. New records - defined by the material (chemical formula) and a related publication DOI -- can be registered by external users via a simple web form.}
 \label{fig:taxonomy}
\end{figure*}

Experimental 2D materials science is a diverse, cross-disciplinary field that encompasses a wide range of synthesis and measurement techniques. To promote a coherent and consistent description of experimental 2D materials data, X2DB uses a high-level taxonomy specifically tailored to atomically thin samples. The taxonomy organizes information into conceptual categories covering crystal structure, sample morphology, synthesis, characterization methods, and measured properties, rather than individual measurement datasets. A schematic overview is provided in Figure \ref{fig:taxonomy}. 

The taxonomy's primary aim is to standardize the representation of experimentally realized 2D materials and to establish a controlled vocabulary for their description. It can be seen as a compact parameterization of "what matters experimentally for 2D materials", designed to support statistical inference across the literature.

In addition to the main categories shown in Figure \ref{fig:taxonomy}, it includes additional sub-categories and predefined labels, e.g. for the most common substrates, synthesis techniques, and characterization methods. A complete description of the taxonomy is provided in the SI. The selection of (sub)categories and labels used in the taxonomy, is based on the 221 publications analyzed at the final stage of the literature mining. While the predefined labels of the taxonomy are often sufficient to describe all relevant aspects of synthesized 2D materials, they are obviously not exhaustive; X2DB therefore also accommodates any user-specified text within the relevant categories. Below we provide an overview of the main categories of the taxonomy. 


\subsection{Structure and Composition}
This category describes the pristine material using standard crystallographic terminology, including its chemical formula, structural phase, and crystal symmetries of the as-grown/as-exfoliated 2D sample. In addition to the conventional 3D space group, the taxonomy also includes the 2D crystal system describing the symmetry of the in-plane lattice (square, rectangular, hexagonal, oblique). This is particularly relevant for ultrathin materials for which the space group may be difficult to determine or not even well-defined due to limited out-of-plane periodicity. Certain classes of 2D materials, in particular MXenes, may contain unresolved mixed/non-stoichiometric surface termination groups (e.g. O, OH, F) resulting from the chemical synthesis. These termination groups can strongly influence the properties of the materials. It is possible to specify such termination groups alongside the chemical formula of the bare, unterminated 2D material. For 2D materials produced by exfoliation technique, the chemical formula and space group of the parent bulk material can be specified separately. 
 
In addition to the symmetry and crystal phase, a number of \emph{structural classifier} keywords can be used to describe the 2D material in greater detail. These include (but are not limited to) \emph{vdW-layered material} and \emph{non-vdW material}, \emph{multiphase} (denoting samples containing distinct regions of different structural phases\cite{feng2021lattice}), \emph{defect-derived} (for crystal structures that are obtained from an ideal parent crystal by introducing defects such as vacancies\cite{van20242d}), \emph{substrate-bonded} (referring to cases where the 2D material is chemically bonded to the substrate, such as silicene on Ag(111)\cite{feng2012evidence}), \emph{intercalated} and \emph{self-intercalated} (for vdW structures with hetero or native atoms in the vdW gaps\cite{zhao2020engineering,karthikeyan2019transition}).

\subsection{Sample Morphology}
The qualitative material features are complemented by quantitative attributes that describe the sample morphology. These comprise the sample thickness (categorized as "$<$1 nm", "1-3 nm", "3-10 nm", "$>$10 nm"), the number of layers - or, for non-layered materials, the number of out-of-plane unit cells - and the lateral flake size. Thickness and lateral size refer to the thinnest samples obtained.

\subsection{Synthesis Methods}
Synthesis methods are classified into bottom-up growth and top-down exfoliation. The bottom-up category is further divided into bulk growth methods, which produce macroscopic crystals or powders for subsequent exfoliation, and direct 2D growth methods, which directly deposit or synthesize the material in atomically thin form. The top-down category encompasses the most widely used exfoliation techniques including dry and liquid-phase mechanical exfoliation, selective etching, and intercalation. 

\subsection{Substrates}
Substrates play a crucial role for both the growth and properties of 2D materials. The taxonomy distinguishes between growth substrates, used for direct synthesis, e.g. chemical vapor deposition (CVD) or molecular beam epitaxy (MBE), and transfer substrates used for characterization and/or device integration. Use of metallic growth substrates often lead to strongly bonded, non-transferable 2D materials (in such cases, it may be more appropriate to use the term adsorption layer rather than a 2D material). In addition to the direct chemical interaction, the substrate also influences electronic properties of the 2D material via non-local dielectric effects, which modify the (screened) Coulomb interactions in the 2D material leading to renormalization of band structures, optical excitations, and transport properties\cite{raja2019dielectric,thygesen2017calculating,radisavljevic2013mobility}. Consequently, not only the chemical inertness and structural/electronic uniformity of the substrate is of importance, but also its dielectric properties.

\subsection{Characterization Methods}
The characterization methods included in the taxonomy cover the most widely used techniques in the field, including scanning and transmission electron microscopy, photoelectron spectroscopies, scanning probe techniques, X-ray absorption and diffraction methods, optical imaging and spectroscopy (including Raman and second harmonic generation), magnetic resonance techniques, and electrical measurements. A free-text field is available for providing more details and/or characterization techniques not covered by the pre-defined descriptors.

\subsection{Properties and Applications}
This category is used to describe the measured properties and/or explored applications reported in the associated publication. The taxonomy provides a set of predefined labels covering the most central topical areas in 2D materials research, such as linear and nonlinear optics, phase transitions, transport regimes, magnetic order, optoelectronics, spintronics, batteries, and (electro)catalysis. Although extensive, the list is not exhaustive and is intentionally kept relatively coarse-grained. A free-text field is available for more detailed descriptions, including properties or applications not covered by the predefined labels, as well as for reporting quantitative measurement data.


\begin{figure*}
 \centering
 \includegraphics[height=7cm]{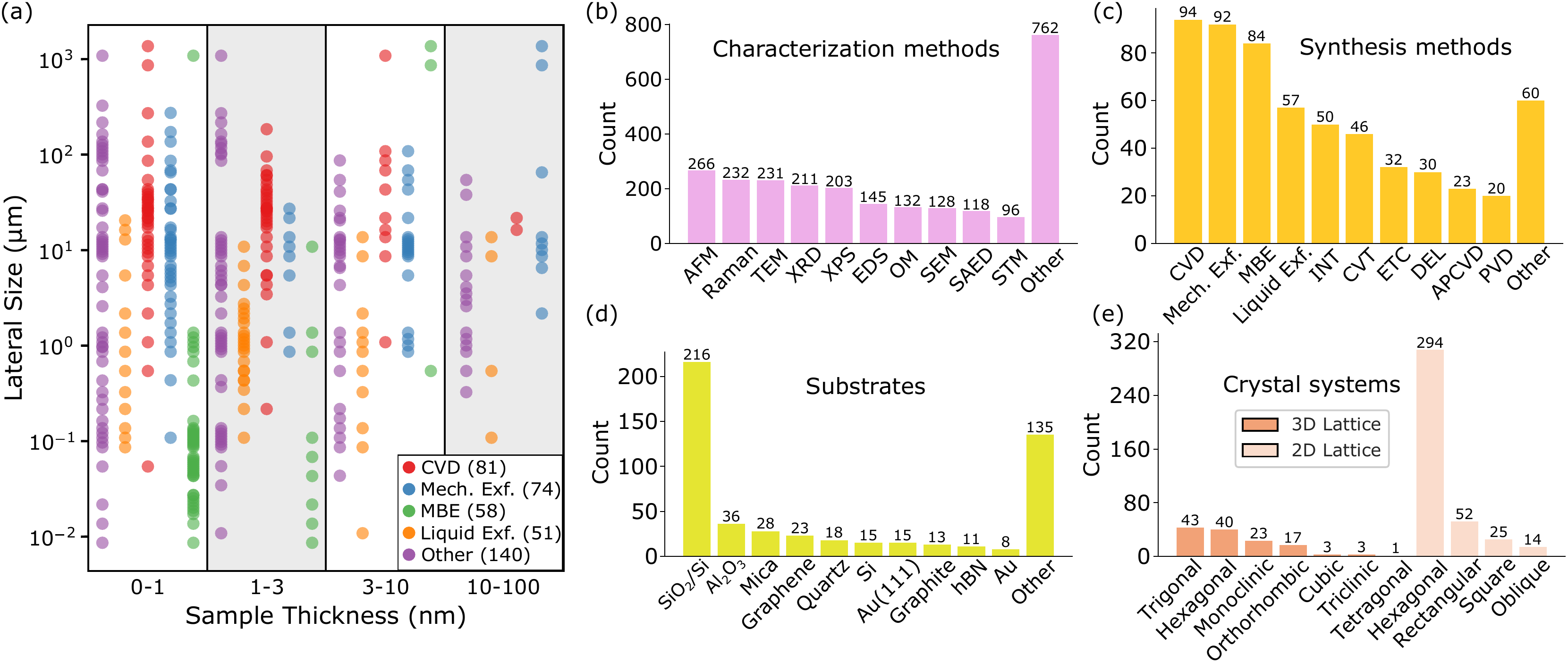}
 \caption{\textbf{Statistical Overview of Selected Data from X2DB}. (a) Morphological parameters (lateral size and thickness) of experimentally realized 2D materials as a function of synthesis method. (b-e) Distributions of the most frequently reported characterization methods, synthesis methods, crystal systems (for both the 2D in-plane lattice and the 3D lattice), and substrates used for growth or transfer. Each plot highlights the top 10 most common reports, while all other, less frequent entries are grouped in a single combined bar to provide a comprehensive overview.}
 \label{fgr:example2col}
\end{figure*} 

\begin{figure*}
 \centering
 \includegraphics[height=11cm]{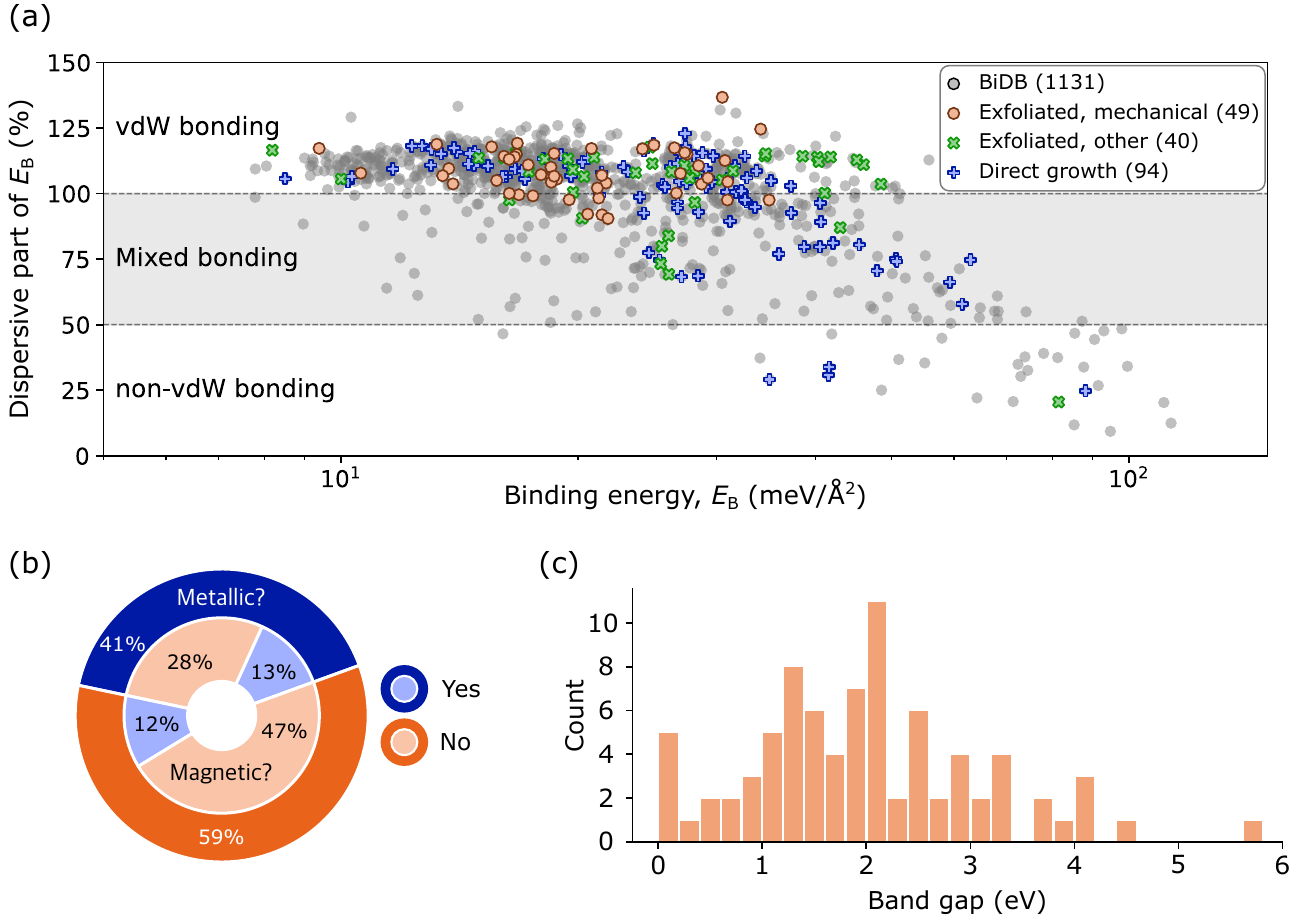}
 \caption{\textbf{Computational Properties of Experimentally Realized 2D Materials.} All colored data refers to materials from X2DB that have been produced in few-layer form with thickness below 10 nm and have been linked to a monolayer in C2DB with high confidence level. (a) Interlayer binding energies calculated with the PBE-D3 method. The contribution of the dispersive D3-term to the binding energy is plotted as function of the total binding energy, $E_\mathrm{B}$. Grey symbols show all materials in the computational bilayer database BiDB. Materials in X2DB are highlighted depending on whether they were mechanically exfoliated (orange), exfoliated bu other techniques (green), or directly grown (blue). The type of interlayer bonding can be classified according to whether it is dominated by dispersive forces (vdW bonding), covalent or ionic bonding (non-vdW bonding), or a mixture of both (mixed bonding regime). (b) Distribution of the calculated electronic and magnetic properties of the monolayer form of the materials in X2DB. The outer ring shows the fraction of metallic vs. non-metallic monolayers. The inner ring shows the fraction of magnetic vs. non-magnetic materials further separated into metallic and non-metallic categories.  (c) The band gap distribution for non-metallic materials, as calculated with the HSE06 xc-functional. 
}
 \label{fgr:compprop}
\end{figure*}


\subsection{Linking to Computational Data}
A key feature of X2DB, is its integration with computational databases. When uploading an entry, users may provide a unique identifier (UID) for a corresponding structure in the C2DB monolayer database and assign the match a confidence level from 1-3. Entries that are matched with a C2DB monolayer structure upon registration, will contain a link to the relevant monolayer in C2DB from its X2DB page. In addition, for matches with high confidence level,
links to the corresponding homobilayer structures (all stable stacking configurations) in the computational bilayer database, BiDB\cite{Pakdel2023EmergentStacking}, will be generated automatically and inserted on the X2DB page. We note that the interlayer binding energies available in BiDB can be regarded as a proxy for the exfoliation energy -- an approach justified by the fact that a bilayer's binding energy is a well-approximated measure (within 10\%) for the exfoliation energy of a thick slab.\cite{jung2018rigorous} Moreover, by comparing the  electronic band structures of the monolayer and homobilayers one can get an idea of the strength and importance of interlayer hybridization upon stacking.  

Similarly, entries that hold a space group for the bulk form of the 2D material, are automatically linked to the relevant bulk crystal structure in the computational database CrystalBank. This makes it possible to explore the detailed crystal structure and the electronic band structure of the bulk compound.

The C2DB, BiDB, and CrystalBank databases are internally consistent: All crystal structures and properties are computed using the same electronic structure code (GPAW\cite{Mortensen2024GPAW:Calculations}) and numerical settings. This uniformity allows direct comparison of properties -- such as structural parameters, magnetic moments, and band gaps -- across monolayers, bilayers, and bulk structures. Had the calculations been performed with different codes, methods, or parameter settings, any variations in the calculated properties could reflect numerical discrepancies rather than intrinsic physical differences.

\section{The Database}
X2DB organizes information in \emph{entries}, each defined by the chemical formula of a material and the DOI of a publication reporting an experimental study of that material in few-layer form. The information contained in an entry is described using the 2D materials taxonomy introduced above. While the taxonomy covers both van der Waals layered crystals and compounds without a naturally layered topology, non-stoichiometric compounds, heterostructures, and twisted moiré structures are currently not supported.

New entries can be uploaded to X2DB by external users via a simple web form. The only requirement is that the user possesses an ORCiD. In particular, they need not be an author of the associated publication. Because each entry represents a material–publication pair, multiple materials can be linked to the same publication, and multiple publications can be linked to the same material. Currently, most materials are associated with only a single or a few publications. However, a popular material like MoS$_2$, has been registered in more than 20 entries, each with a different publication. Such diversity in publication space is highly useful as it can provide an overview of different employed synthesis methods and resulting sample morphologies as well the range of property measurements and applications explored for a given 2D material.  

As an example of the type of information available in X2DB, Fig. \ref{fgr:example2col} presents some statistical overviews of selected data sets. Fig. \ref{fgr:example2col}(a) shows the lateral size versus thickness of 2D samples obtained by different synthesis/exfoliation methods, where each point represents one of the 470 entries in X2DB. It can be seen that molecular beam epitaxy (MBE) typically produces thin (mostly monolayer) samples of small lateral dimensions ($<1\mu\mathrm{m}$). In contrast, chemical vapor deposition (CVD) yields samples with a broader thickness range and significantly larger lateral sizes ($10-100\mu\mathrm{m}$). Dry mechanical exfoliation methods (e.g. Scotch tape/stamping) produce samples with thickness and lateral size distributions similar to those obtained by CVD, whereas liquid-phase exfoliation techniques (e.g., ultrasonication or high-shear methods) generally yield slightly thicker samples with smaller lateral sizes ($0.1-10\mu\mathrm{m}$). 

Figures~\ref{fgr:example2col}(b–d) show the distributions of registered materials over characterization methods, synthesis routes, and substrates used for growth or transfer. In each case, the ten most frequent categories are shown explicitly, while the remainder are grouped under “Others”. While there is a clear predominance of SiO$_2$	substrates, no single synthesis or characterization method stands out as particularly dominant. Instead, the distributions are relatively flat with long tails (note the sizable “Others” category), highlighting the large diversity of techniques employed in the field. Panel (e) shows the distribution of crystal systems (for both the 2D in-plane lattice and the 3D lattice) of the registered 2D materials. There is a pronounced predominance of hexagonal crystal structures, consistent with the close-packed nature of most 2D materials. We note that there are fewer registrations of the 3D crystal symmetry. This is because determination of the 3D space group, or even just the 3D crystal system, requires samples of a certain thickness in order for the out-of-plane periodicity to be well defined and measurable. For very thin samples, in particular monolayers and bilayers, the concept is not even meaningful (see also the discussion in Sec.~\ref{sec:class} on space groups versus layer groups).
All abbreviations used in Figure \ref{fgr:example2col} are defined in Tables S1 and S2 in the SI.  

\section{Integration of Computational Data}
The integration of computational data with X2DB enables direct comparison between measured and calculated properties, as well as exploration of predicted properties beyond those accessible experimentally. Below we give a few illustrative examples.

Figure \ref{fgr:compprop}(a) shows the  calculated exfoliation energy (more precisely the binding energy of the homobilayer in its most stable stacking configuration) plotted along the $x$-axis for all the materials in X2DB that have been matched with a monolayer in C2DB with high confidence level and have been realized with minimum thickness below 10 nm. The grey symbols represent all the homobilayers in the computational homobilayer database BiDB\cite{pakdel2024high}. We note in passing that BiDB contains more than 3,000 homobilayers constructed by stacking 1000 of the most stable monolayers from C2DB in all possible lattice matched, i.e. non-moiré, stacking configurations. In Figure \ref{fgr:compprop}(a), the materials from X2DB have been grouped according to synthesis method: Orange symbols denote materials with at least one entry in X2DB reporting synthesis by direct, i.e. non-assisted, mechanical exfoliation (e.g. Scotch-tape method); green symbols represent materials with at least one entry reporting exfoliation by some other techniques (e.g. ultrasonication, electrochemical intercalation, or selective etching); blue symbols represent bottom-up grown materials (e.g. CVD, MBE). Note that the division is somewhat ambiguous as many materials can be produced by various methods. 
However, the results may still be used to explore limits for exfoliation. In particular, it is clear that except for a couple of outliers, all mechanically exfoliated materials have binding energy below 40 meV/\AA$^2$.

The binding energies in Figure \ref{fgr:compprop}(a) were calculated using the PBE-D3 functional, where the D3 correction adds an empirical description of long-range van der Waals (dispersion) forces that the standard PBE functional otherwise misses.\cite{grimme2011effect} The values on the $y$-axis of Figure \ref{fgr:compprop}(a) show the relative contribution of the D3 term to the total interlayer binding energy.   On this basis, three different bonding regimes can be identified: The vdW regime, where the PBE (i.e. non-vdW) contribution to the binding energy is negative due to Pauli repulsion. The non-vdW regime, where the PBE contribution to $E_{\mathrm{B}}$ dominates the dispersive D3 contribution, and the mixed bonding regime, where PBE contributes to the binding but does not exceed the D3 contribution.

While most of the mechanically exfoliated materials in Figure \ref{fgr:compprop}(a) fall within the vdW bonding region, several materials obtained with other exfoliation methods lie in the mixed bonding regime. The 2D electride Ca$_2$N is a notable outlier.\cite{druffel2016experimental} Located in the non-vdW region, Ca$_2$N exhibits a large binding energy of 80 meV/\AA$^2$, yet it has been exfoliated using ultrasonication. This unique behavior is attributed to electrostatic interactions between the [Ca$_2$N]$^+$ layers and the interstitial electron gas, which are fundamentally different from the weaker vdW forces governing bonding in the dispersive regime and the stronger and more directional covalent bonds. Among the directly grown 2D materials, several are located in the mixed and non-vdW bonding regions with binding energies reaching 90 meV/\AA$^2$.   

Figure \ref{fgr:compprop}(b) shows an overview of the calculated electronic type (metallic or non-metallic) and magnetic state of the materials in X2DB. Only materials with a high C2DB confidence level are included. The results refer to the monolayer form of the materials. It can be seen from the outer ring of the diagram that 59\% of the realized 2D materials are predicted by DFT to be semiconductors or insulators while 41\% are predicted to be metals.  The inner shell reveals that 25\% of the materials are predicted to favor a magnetic over a non-magnetic ground state. Moreover, there is a relatively higher fraction of magnetic materials among the metals than among the non-metals. 

We emphasize that the DFT calculations presented in Figure \ref{fgr:compprop}(b) do not capture strong electronic correlations, including possible Mott insulating states. Consequently, the computed fraction of metallic materials may be higher than what is observed experimentally. A recent assessment of electronic and magnetic properties of more than 600 monolayer materials containing $3d$ transition metals, showed that introducing a Hubbard U-term in the DFT Hamiltonian, to improve the description of localized states, resulted a metal-to-insulator transition in 21\% of the materials. In contrast, the magnetic ground state was largely unaffected by the U-term.\cite{pakdel2025effect}

Finally, Figure \ref{fgr:compprop}(c) shows the distribution of band gaps calculated with the HSE06 hybrid functional\cite{krukau2006influence} for the non-metallic monolayers in X2DB. Again, only materials with a high C2DB confidence level are shown. The band gaps span from 0 to 5.7 eV (with hBN having the largest gap). In general, it is expected that bilayer and multilayer structures will have smaller band gaps than the monolayers (shown here), due to interlayer hybridization and increased dielectric screening -- both of which tend to reduce the fundamental gap. On the other hand, the HSE06 functional typically underestimates fundamental band gaps compared to experiments and more accurate many-body methods. In fact, the HSE06 was found to underestimate the G$_0$W$_0$ band gap by around 20\% for a set of 250 2D semiconductors\cite{Haastrup2018TheCrystals}.

\section{Classification of 2D Materials}\label{sec:class}
\begin{figure*}
\centering
 \includegraphics[height=11cm]{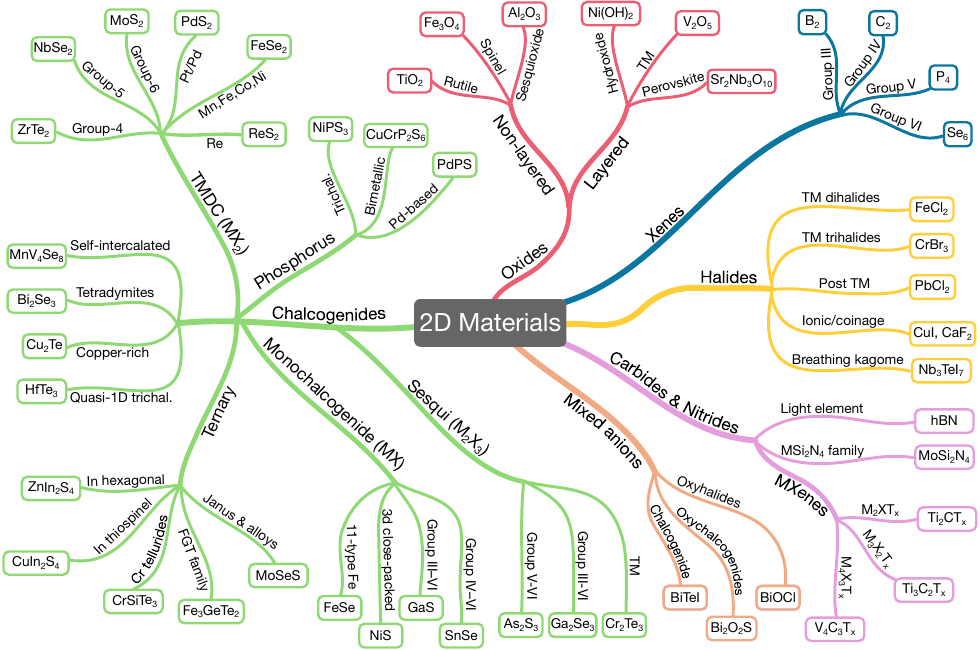}
 \caption{\textbf{Hierarchical classification tree of experimentally realized 2D materials.} The main classes are distinguished by different colors (consistent with Table I) and branch out into subclasses and terminal classes. The hierarchy of classes is determined primarily according to anion-type, chemical composition/stoichiometry, and crystal structure, respectively. One representative member from each terminal class is shown at the end of the branches.}
 \label{fig:class}
\end{figure*}

To provide a systematic overview of all experimentally realized 2D materials, we have constructed a hierarchical classification based on the curated data in X2DB. The classification is listed in Table I and shown graphically in Fig.~\ref{fig:class}. The main classes are defined primarily by the anion type and comprise the historically important elemental 2D materials (Xenes), the chalcogenides, the halides, the oxides, the carbides and nitrides (including MXenes), and mixed-anion compounds. In Fig.~\ref{fig:class} and Table I, the main classes are distinguished by color. Subclasses are mainly defined by chemical stoichiometry and/or cation group, and the final end-classes typically contain 3–10 members with similar chemical composition, structural type, and basic electronic/chemical properties.

Beyond providing a compact overview, this classification serves as an organisational backbone for the X2DB data. It enables comparison of trends within and across chemically related families (e.g.\ band gaps, stability, or magnetic order within a given anion class), highlights sparsely populated or entirely missing regions of 2D materials space, and thus points to promising directions for targeted synthesis. The hierarchy also defines natural ``families'' that can be used to construct balanced benchmark sets for machine-learning models, to design property-screening studies focused on specific application domains (such as catalysis or optoelectronics), and to consistently place newly reported 2D materials into the existing landscape.

The right column in Table I provides a brief summary of the main structural and electronic characteristics of the members of each class, including their typical crystal symmetry groups and main application area. We emphasize that space groups (SGs) should be used exclusively to describe a bulk/multilayer form of a 2D material where the out-of-plane periodicity is well defined. In contrast, monolayer structures, which lack out-of-plane periodicity, are better described by their layer group (LG).\cite{fu2024symmetry} The same monolayer can be stacked in different ways to produce bulk or multilayer structures with different SGs. To avoid this ambiguity, we consistently use LGs for monolayers and SGs for bulk/multilayer structures.


\section{Discussion}
The open X2DB database with its curated, taxonomy-based data set and field-wide classification of experimentally realized 2D materials, provides a structured, global overview of the rich compound space that underpins the rapidly evolving field of 2D materials. Its tight integration of experimental and computational data enables seamless navigation between real-world measurements and theoretical predictions opening new opportunities to interrogate the factors governing stability and synthesizability, perform statistical inference across the literature, and guide the targeted discovery of novel 2D compounds. X2DB can also be used to identify “holes” in the current 2D materials map — arising, for example, from biases in the choice of anions, metals, or synthesis strategies — and to infer promising synthesis routes and growth substrates for specific compositions and material families.

At the time of writing, the X2DB contains 370 unique compounds, a number that is likely close to the full set of 2D materials that have been experimentally realized to date. Of these, 210 materials have been matched with high confidence to their monolayer, homobilayer, and bulk counterparts in mutually consistent first-principles databases.
 However, the purpose of X2DB extends far beyond serving as a showcase of previously synthesized 2D materials linked to computational data. By collecting multiple experimental records for the same compound, the database enables systematic exploration of how sample morphology and quality vary with synthesis approach, substrate, and related parameters, and it supports comprehensive overviews of the characterization techniques, measured physical properties, and reported applications for a given material of interest.
 Likewise, X2DB can be used to identify publications reporting specific types of measurements -- such as Raman spectra, charge density waves, excitons, magnons, etc. -- across different materials.

While experimental data sets can be contributed through a simple web form, only data related to published work can currently be be registered. This requirement ensures the highest standards of reliability, traceability, and documentation of the data. Future developments may allow the inclusion of unpublished results, should the need arise. 

The continued growth and usefulness of X2DB will rely on active community engagement.  By contributing their experimental data and publications, researchers not only increase the visibility and FAIRness of their work, but also enrich it by enabling links to complementary computational data. Beyond individual benefits, such data contributions are essential for advancing X2DB to a dynamic community resource that evolves continuously and provides lasting value to both experimentalists and theorists across the 2D materials field. 

\section{Data availability}
The open X2DB database is available at https://x2db.fysik.dtu.dk/.

\clearpage

\bibliographystyle{apsrev4-2}
\bibliography{ref_main}

@article{He2024AdvancesConversion,
  title={Advances and challenges in MXene-based electrocatalysts: unlocking the potential for sustainable energy conversion},
  author={He, Lei and Zhuang, Haizheng and Fan, Qi and Yu, Ping and Wang, Shengchao and Pang, Yifan and Chen, Ke and Liang, Kun},
  journal={Materials Horizons},
  volume={11},
  number={18},
  pages={4239--4255},
  year={2024},
  publisher={Royal Society of Chemistry}
}

@incollection{Anasori20172DStorage,
  title={2D metal carbides and nitrides (MXenes) for energy storage},
  author={Anasori, Babak and Lukatskaya, Maria R and Gogotsi, Yury},
  booktitle={MXenes},
  pages={677--722},
  year={2023},
  publisher={Jenny Stanford Publishing}
}

@article{doi:10.1021/acsphotonics.5c00353,
author = {de Abajo, F. Javier GarcÃ­a and Basov, D. N. and Koppens, Frank H. L. and Orsini, Lorenzo and others
    },
title = {Roadmap for Photonics with 2D Materials},
journal = {ACS Photonics},
volume = {12},
number = {8},
pages = {3961-4095},
year = {2025},
doi = {10.1021/acsphotonics.5c00353},
URL = { https://doi.org/10.1021/acsphotonics.5c00353
}
}

@article{Song2018Two-DimensionalApplications,
    title = {{Two-Dimensional Materials for Thermal Management Applications}},
    year = {2018},
    journal = {Joule},
    author = {Song, Houfu and Liu, Jiaman and Liu, Bilu and Wu, Junqiao and Cheng, Hui-Ming and Kang, Feiyu},
    number = {3},
    month = {3},
    pages = {442--463},
    volume = {2},
    doi = {10.1016/j.joule.2018.01.006},
    issn = {25424351}
}

@article{liu20192d,
  title={2D materials for quantum information science},
  author={Liu, Xiaolong and Hersam, Mark C},
  journal={Nature Reviews Materials},
  volume={4},
  number={10},
  pages={669--684},
  year={2019},
  publisher={Nature Publishing Group UK London}
}

@article{grimme2011effect,
  title={Effect of the damping function in dispersion corrected density functional theory},
  author={Grimme, Stefan and Ehrlich, Stephan and Goerigk, Lars},
  journal={Journal of Computational Chemistry},
  volume={32},
  number={7},
  pages={1456--1465},
  year={2011},
  publisher={Wiley Online Library}
}

@article{krukau2006influence,
  title={Influence of the exchange screening parameter on the performance of screened hybrid functionals},
  author={Krukau, Aliaksandr V and Vydrov, Oleg A and Izmaylov, Artur F and Scuseria, Gustavo E},
  journal={The Journal of Chemical Physics},
  volume={125},
  number={22},
  year={2006},
  publisher={AIP Publishing}
}

@article{pakdel2025effect,
  title={Effect of Hubbard U-corrections on the electronic and magnetic properties of 2D materials: a high-throughput study},
  author={Pakdel, Sahar and Olsen, Thomas and Thygesen, Kristian S},
  journal={npj Computational Materials},
  volume={11},
  number={1},
  pages={18},
  year={2025},
  publisher={Nature Publishing Group UK London}
}

@article{Song2010LargeLayers,
    title = {{Large Scale Growth and Characterization of Atomic Hexagonal Boron Nitride Layers}},
    year = {2010},
    journal = {Nano Letters},
    author = {Song, Li and Ci, Lijie and Lu, Hao and Sorokin, Pavel B. and Jin, Chuanhong and Ni, Jie and Kvashnin, Alexander G. and Kvashnin, Dmitry G. and Lou, Jun and Yakobson, Boris I. and Ajayan, Pulickel M.},
    number = {8},
    month = {8},
    pages = {3209--3215},
    volume = {10},
    doi = {10.1021/nl1022139},
    issn = {1530-6984}
}

@article{Wang2012ElectronicsDichalcogenides,
    title = {{Electronics and optoelectronics of two-dimensional transition metal dichalcogenides}},
    year = {2012},
    journal = {Nature Nanotechnology},
    author = {Wang, Qing Hua and Kalantar-Zadeh, Kourosh and Kis, Andras and Coleman, Jonathan N. and Strano, Michael S.},
    number = {11},
    month = {11},
    pages = {699--712},
    volume = {7},
    doi = {10.1038/nnano.2012.193},
    issn = {1748-3387}
}

@article{zhao20212d,
  title={2D metallic transition-metal dichalcogenides: structures, synthesis, properties, and applications},
  author={Zhao, Bei and Shen, Dingyi and Zhang, Zucheng and Lu, Ping and Hossain, Mongur and Li, Jia and Li, Bo and Duan, Xidong},
  journal={Advanced Functional Materials},
  volume={31},
  number={48},
  pages={2105132},
  year={2021},
  publisher={Wiley Online Library}
}

@article{Novoselov2004ElectricFilms,
    title = {{Electric Field Effect in Atomically Thin Carbon Films}},
    year = {2004},
    journal = {Science},
    author = {Novoselov, K. S. and Geim, A. K. and Morozov, S. V. and Jiang, D. and Zhang, Y. and Dubonos, S. V. and Grigorieva, I. V. and Firsov, A. A.},
    number = {5696},
    month = {10},
    pages = {666--669},
    volume = {306},
    doi = {10.1126/science.1102896},
    issn = {0036-8075}
}

@article{wang2023towards,
  title={Towards two-dimensional van der Waals ferroelectrics},
  author={Wang, Chuanshou and You, Lu and Cobden, David and Wang, Junling},
  journal={Nature Materials},
  volume={22},
  number={5},
  pages={542--552},
  year={2023},
  publisher={Nature Publishing Group UK London}
}

@article{wang2017high,
  title={High-quality monolayer superconductor NbSe2 grown by chemical vapour deposition},
  author={Wang, Hong and Huang, Xiangwei and Lin, Junhao and Cui, Jian and Chen, Yu and Zhu, Chao and Liu, Fucai and Zeng, Qingsheng and Zhou, Jiadong and Yu, Peng and others},
  journal={Nature Communications},
  volume={8},
  number={1},
  pages={394},
  year={2017},
  publisher={Nature Publishing Group UK London}
}

@article{xu2018electrically,
  title={Electrically switchable Berry curvature dipole in the monolayer topological insulator WTe2},
  author={Xu, Su-Yang and Ma, Qiong and Shen, Huitao and Fatemi, Valla and Wu, Sanfeng and Chang, Tay-Rong and Chang, Guoqing and Valdivia, Andr{\'e}s M Mier and Chan, Ching-Kit and Gibson, Quinn D and others},
  journal={Nature Physics},
  volume={14},
  number={9},
  pages={900--906},
  year={2018},
  publisher={Nature Publishing Group UK London}
}

@article{zagorac2019recent,
  title={Recent developments in the Inorganic Crystal Structure Database: theoretical crystal structure data and related features},
  author={Zagorac, Dejan and M{\"u}ller, H and Ruehl, S and Zagorac, J and Rehme, Silke},
  journal={Applied Crystallography},
  volume={52},
  number={5},
  pages={918--925},
  year={2019},
  publisher={International Union of Crystallography}
}

@article{jain2013commentary,
  title={Commentary: The Materials Project: A materials genome approach to accelerating materials innovation},
  author={Jain, Anubhav and Ong, Shyue Ping and Hautier, Geoffroy and Chen, Wei and Richards, William Davidson and Dacek, Stephen and Cholia, Shreyas and Gunter, Dan and Skinner, David and Ceder, Gerbrand and others},
  journal={APL Materials},
  volume={1},
  number={1},
  year={2013},
  publisher={AIP Publishing}
}

@article{Haastrup2018TheCrystals,
    title = {{The Computational 2D Materials Database: high-throughput modeling and discovery of atomically thin crystals}},
    year = {2018},
    journal = {2D Materials},
    author = {Haastrup, Sten and Strange, Mikkel and Pandey, Mohnish and Deilmann, Thorsten and Schmidt, Per S and Hinsche, Nicki F and Gjerding, Morten N and Torelli, Daniele and Larsen, Peter M and Riis-Jensen, Anders C and Gath, Jakob and Jacobsen, Karsten W and J{\o}rgen Mortensen, Jens and Olsen, Thomas and Thygesen, Kristian S},
    number = {4},
    month = {9},
    pages = {042002},
    volume = {5},
    doi = {10.1088/2053-1583/aacfc1},
    issn = {2053-1583}
}

@article{campi2023expansion,
  title={Expansion of the materials cloud 2D database},
  author={Campi, Davide and Mounet, Nicolas and Gibertini, Marco and Pizzi, Giovanni and Marzari, Nicola},
  journal={ACS Nano},
  volume={17},
  number={12},
  pages={11268--11278},
  year={2023},
  publisher={ACS Publications}
}

@article{zhou20192dmatpedia,
  title={2DMatPedia, an open computational database of two-dimensional materials from top-down and bottom-up approaches},
  author={Zhou, Jun and Shen, Lei and Costa, Miguel Dias and Persson, Kristin A and Ong, Shyue Ping and Huck, Patrick and Lu, Yunhao and Ma, Xiaoyang and Chen, Yiming and Tang, Hanmei and others},
  journal={Scientific Data},
  volume={6},
  number={1},
  pages={86},
  year={2019},
  publisher={Nature Publishing Group UK London}
}

@article{lebegue2013two,
  title={Two-dimensional materials from data filtering and ab initio calculations},
  author={Leb{\`e}gue, S{\'e}bastien and Bj{\"o}rkman, Torbj{\"o}rn and Klintenberg, Mattias and Nieminen, Risto M and Eriksson, Olle},
  journal={Physical Review X},
  volume={3},
  number={3},
  pages={031002},
  year={2013},
  publisher={APS}
}

@article{cheon2017data,
  title={Data mining for new two-and one-dimensional weakly bonded solids and lattice-commensurate heterostructures},
  author={Cheon, Gowoon and Duerloo, Karel-Alexander N and Sendek, Austin D and Porter, Chase and Chen, Yuan and Reed, Evan J},
  journal={Nano Letters},
  volume={17},
  number={3},
  pages={1915--1923},
  year={2017},
  publisher={ACS Publications}
}

@article{ashton2017topology,
  title={Topology-scaling identification of layered solids and stable exfoliated 2D materials},
  author={Ashton, Michael and Paul, Joshua and Sinnott, Susan B and Hennig, Richard G},
  journal={Physical Review Letters},
  volume={118},
  number={10},
  pages={106101},
  year={2017},
  publisher={APS}
}

@article{mounet2018two,
  title={Two-dimensional materials from high-throughput computational exfoliation of experimentally known compounds},
  author={Mounet, Nicolas and Gibertini, Marco and Schwaller, Philippe and Campi, Davide and Merkys, Andrius and Marrazzo, Antimo and Sohier, Thibault and Castelli, Ivano Eligio and Cepellotti, Andrea and Pizzi, Giovanni and others},
  journal={Nature Nanotechnology},
  volume={13},
  number={3},
  pages={246--252},
  year={2018},
  publisher={Nature Publishing Group UK London}
}

@article{bjork2024two,
  title={Two-dimensional materials by large-scale computations and chemical exfoliation of layered solids},
  author={Bj{\"o}rk, Jonas and Zhou, Jie and Persson, Per O{\AA} and Rosen, Johanna},
  journal={Science},
  volume={383},
  number={6688},
  pages={1210--1215},
  year={2024},
  publisher={American Association for the Advancement of Science}
}

@article{feng2021lattice,
  title={Lattice-matched metal--semiconductor heterointerface in monolayer Cu2Te},
  author={Feng, Jingqi and Gao, Huiying and Li, Tian and Tan, Xin and Xu, Peng and Li, Menglei and He, Lin and Ma, Donglin},
  journal={ACS Nano},
  volume={15},
  number={2},
  pages={3415--3422},
  year={2021},
  publisher={ACS Publications}
}

@article{van20242d,
  title={2D Vanadium Sulfides: Synthesis, Atomic Structure Engineering, and Charge Density Waves},
  author={van Efferen, Camiel and Hall, Joshua and Atodiresei, Nicolae and Boix, Virginia and Safeer, Affan and Wekking, Tobias and Vinogradov, Nikolay A and Preobrajenski, Alexei B and Knudsen, Jan and Fischer, Jeison and others},
  journal={ACS Nano},
  volume={18},
  number={22},
  pages={14161--14175},
  year={2024},
  publisher={ACS Publications}
}

@article{feng2012evidence,
  title={Evidence of silicene in honeycomb structures of silicon on Ag (111)},
  author={Feng, Baojie and Ding, Zijing and Meng, Sheng and Yao, Yugui and He, Xiaoyue and Cheng, Peng and Chen, Lan and Wu, Kehui},
  journal={Nano Letters},
  volume={12},
  number={7},
  pages={3507--3511},
  year={2012},
  publisher={ACS Publications}
}

@article{zhao2020engineering,
  title={Engineering covalently bonded 2D layered materials by self-intercalation},
  author={Zhao, Xiaoxu and Song, Peng and Wang, Chengcai and Riis-Jensen, Anders C and Fu, Wei and Deng, Ya and Wan, Dongyang and Kang, Lixing and Ning, Shoucong and Dan, Jiadong and others},
  journal={Nature},
  volume={581},
  number={7807},
  pages={171--177},
  year={2020},
  publisher={Nature Publishing Group UK London}
}

@article{gibertini2019magnetic,
  title={Magnetic 2D materials and heterostructures},
  author={Gibertini, Magnetic and Koperski, Maciej and Morpurgo, Alberto F and Novoselov, Konstantin S},
  journal={Nature nanotechnology},
  volume={14},
  number={5},
  pages={408--419},
  year={2019},
  publisher={Nature Publishing Group UK London}
}

@article{americo2025predicting,
  title={Predicting Aqueous and Electrochemical Stability of 2D Materials from Extended Pourbaix Analyses},
  author={Americo, Stefano and Castelli, Ivano E and Thygesen, Kristian Sommer},
  journal={ACS Electrochemistry},
  year={2025},
  publisher={ACS Publications}
}

@article{Lin2023RecentApplications,
    title = {{Recent Advances in 2D Material Theory, Synthesis, Properties, and Applications}},
    year = {2023},
    journal = {ACS Nano},
    author = {Lin, Yu-Chuan and Torsi, Riccardo and Younas, Rehan and Hinkle, Christopher L. and Rigosi, Albert F. and Hill, Heather M. and Zhang, Kunyan and Huang, Shengxi and Shuck, Christopher E. and Chen, Chen and Lin, Yu-Hsiu and Maldonado-Lopez, Daniel and Mendoza-Cortes, Jose L. and Ferrier, John and Kar, Swastik and Nayir, Nadire and Rajabpour, Siavash and van Duin, Adri C. T. and Liu, Xiwen and Jariwala, Deep and Jiang, Jie and Shi, Jian and Mortelmans, Wouter and Jaramillo, Rafael and Lopes, Joao Marcelo J. and Engel-Herbert, Roman and Trofe, Anthony and Ignatova, Tetyana and Lee, Seng Huat and Mao, Zhiqiang and Damian, Leticia and Wang, Yuanxi and Steves, Megan A. and Knappenberger, Kenneth L. and Wang, Zhengtianye and Law, Stephanie and Bepete, George and Zhou, Da and Lin, Jiang-Xiazi and Scheurer, Mathias S. and Li, Jia and Wang, Pengjie and Yu, Guo and Wu, Sanfeng and Akinwande, Deji and Redwing, Joan M. and Terrones, Mauricio and Robinson, Joshua A.},
    number = {11},
    month = {6},
    pages = {9694--9747},
    volume = {17},
    doi = {10.1021/acsnano.2c12759},
    issn = {1936-0851}
}

@article{ren20252d,
  title={The 2d materials roadmap},
  author={Ren, Wencai and Boggild, Peter and Redwing, Joan M and Novoselov, Konstantin S and Sun, Luzhao and Qi, Yue and Jia, Kaicheng and Liu, Zhongfan and Burton, Oliver and Alexander-Webber, Jack Allen and others},
  journal={2D Materials},
  year={2025}
}

@article{radisavljevic2013mobility,
  title={Mobility engineering and a metal--insulator transition in monolayer MoS2},
  author={Radisavljevic, Branimir and Kis, Andras},
  journal={Nature materials},
  volume={12},
  number={9},
  pages={815--820},
  year={2013},
  publisher={Nature Publishing Group UK London}
}

@article{raja2019dielectric,
  title={Dielectric disorder in two-dimensional materials},
  author={Raja, Archana and Waldecker, Lutz and Zipfel, Jonas and Cho, Yeongsu and Brem, Samuel and Ziegler, Jonas D and Kulig, Marvin and Taniguchi, Takashi and Watanabe, Kenji and Malic, Ermin and others},
  journal={Nature nanotechnology},
  volume={14},
  number={9},
  pages={832--837},
  year={2019},
  publisher={Nature Publishing Group UK London}
}

@article{thygesen2017calculating,
  title={Calculating excitons, plasmons, and quasiparticles in 2D materials and van der Waals heterostructures},
  author={Thygesen, Kristian Sommer},
  journal={2D Materials},
  volume={4},
  number={2},
  pages={022004},
  year={2017},
  publisher={IOP Publishing}
}

@article{karthikeyan2019transition,
  title={Which transition metal atoms can be embedded into two-dimensional molybdenum dichalcogenides and add magnetism?},
  author={Karthikeyan, J and Komsa, Hannu-Pekka and Batzill, Matthias and Krasheninnikov, Arkady V},
  journal={Nano Letters},
  volume={19},
  number={7},
  pages={4581--4587},
  year={2019},
  publisher={ACS Publications}
}

@article{manti2023exploring,
  title={Exploring and machine learning structural instabilities in 2D materials},
  author={Manti, Simone and Svendsen, Mark Kamper and Kn{\o}sgaard, Nikolaj R and Lyngby, Peder M and Thygesen, Kristian S},
  journal={NPJ computational materials},
  volume={9},
  number={1},
  pages={33},
  year={2023},
  publisher={Nature Publishing Group UK London}
}

@article{Ryu2022UnderstandingLearning,
    title = {{Understanding, discovery, and synthesis of 2D materials enabled by machine learning}},
    year = {2022},
    journal = {Chemical Society Reviews},
    author = {Ryu, Byunghoon and Wang, Luqing and Pu, Haihui and Chan, Maria K. Y. and Chen, Junhong},
    number = {6},
    pages = {1899--1925},
    volume = {51},
    doi = {10.1039/D1CS00503K},
    issn = {0306-0012}
}

@article{Choi2022Large-scaleIndustrialization,
    title = {{Large-scale synthesis of graphene and other 2D materials towards industrialization}},
    year = {2022},
    journal = {Nature Communications},
    author = {Choi, Soo Ho and Yun, Seok Joon and Won, Yo Seob and Oh, Chang Seok and Kim, Soo Min and Kim, Ki Kang and Lee, Young Hee},
    number = {1},
    month = {3},
    pages = {1484},
    volume = {13},
    doi = {10.1038/s41467-022-29182-y},
    issn = {2041-1723}
}

@article{Zhao2024ElectrochemicalGraphene,
    title = {{Electrochemical exfoliation of 2D materials beyond graphene}},
    year = {2024},
    journal = {Chemical Society Reviews},
    author = {Zhao, Minghao and Casiraghi, Cinzia and Parvez, Khaled},
    number = {6},
    pages = {3036--3064},
    volume = {53},
    doi = {10.1039/D3CS00815K},
    issn = {0306-0012}
}

@article{Wang2024SynthesisSubstitution,
    title = {{Synthesis and Modulation of Low-Dimensional Transition Metal Chalcogenide Materials via Atomic Substitution}},
    year = {2024},
    journal = {Nano-Micro Letters},
    author = {Wang, Xuan and Chen, Akang and Wu, XinLei and Zhang, Jiatao and Dong, Jichen and Zhang, Leining},
    number = {1},
    month = {12},
    pages = {163},
    volume = {16},
    doi = {10.1007/s40820-024-01378-5},
    issn = {2311-6706}
}

@article{Zhang20222DElectro/Photocatalysis,
    title = {{2D Materials Bridging Experiments and Computations for Electro/Photocatalysis}},
    year = {2022},
    journal = {Advanced Energy Materials},
    author = {Zhang, Xu and Chen, An and Chen, Letian and Zhou, Zhen},
    number = {4},
    month = {1},
    volume = {12},
    doi = {10.1002/aenm.202003841},
    issn = {1614-6832}
}

@article{ma2016metallic,
  title={A metallic mosaic phase and the origin of Mott-insulating state in 1T-TaS2},
  author={Ma, Liguo and Ye, Cun and Yu, Yijun and Lu, Xiu Fang and Niu, Xiaohai and Kim, Sejoong and Feng, Donglai and Tom{\'a}nek, David and Son, Young-Woo and Chen, Xian Hui and others},
  journal={Nature communications},
  volume={7},
  number={1},
  pages={10956},
  year={2016},
  publisher={Nature Publishing Group UK London}
}

@article{pakdel2024high,
  title={High-throughput computational stacking reveals emergent properties in natural van der Waals bilayers},
  author={Pakdel, Sahar and Rasmussen, Asbj{\o}rn and Taghizadeh, Alireza and Kruse, Mads and Olsen, Thomas and Thygesen, Kristian S},
  journal={Nature Communications},
  volume={15},
  number={1},
  pages={932},
  year={2024},
  publisher={Nature Publishing Group UK London}
}

@article{jung2018rigorous,
  title={A rigorous method of calculating exfoliation energies from first principles},
  author={Jung, Jong Hyun and Park, Cheol-Hwan and Ihm, Jisoon},
  journal={Nano letters},
  volume={18},
  number={5},
  pages={2759--2765},
  year={2018},
  publisher={ACS Publications}
}

@article{Mortensen2024GPAW:Calculations,
    title = {{GPAW: An open Python package for electronic structure calculations}},
    year = {2024},
    journal = {The Journal of Chemical Physics},
    author = {Mortensen, Jens Jørgen and Larsen, Ask Hjorth and Kuisma, Mikael and Ivanov, Aleksei V. and Taghizadeh, Alireza and Peterson, Andrew and Haldar, Anubhab and Dohn, Asmus Ougaard and Sch{\"{a}}fer, Christian and J{\'{o}}nsson, Elvar Örn and Hermes, Eric D. and Nilsson, Fredrik Andreas and Kastlunger, Georg and Levi, Gianluca and J{\'{o}}nsson, Hannes and H{\"{a}}kkinen, Hannu and Fojt, Jakub and Kangsabanik, Jiban and S{\o}dequist, Joachim and Lehtom{\"{a}}ki, Jouko and Heske, Julian and Enkovaara, Jussi and Winther, Kirsten Trøstrup and Dulak, Marcin and Melander, Marko M. and Ovesen, Martin and Louhivuori, Martti and Walter, Michael and Gjerding, Morten and Lopez-Acevedo, Olga and Erhart, Paul and Warmbier, Robert and W{\"{u}}rdemann, Rolf and Kaappa, Sami and Latini, Simone and Boland, Tara Maria and Bligaard, Thomas and Skovhus, Thorbjørn and Susi, Toma and Maxson, Tristan and Rossi, Tuomas and Chen, Xi and Schmerwitz, Yorick Leonard A. and Schi{\o}tz, Jakob and Olsen, Thomas and Jacobsen, Karsten Wedel and Thygesen, Kristian Sommer},
    number = {9},
    month = {3},
    volume = {160},
    doi = {10.1063/5.0182685},
    issn = {0021-9606}
}

@article{druffel2016experimental,
  title={Experimental demonstration of an electride as a 2D material},
  author={Druffel, Daniel L and Kuntz, Kaci L and Woomer, Adam H and Alcorn, Francis M and Hu, Jun and Donley, Carrie L and Warren, Scott C},
  journal={Journal of the American Chemical Society},
  volume={138},
  number={49},
  pages={16089--16094},
  year={2016},
  publisher={ACS Publications}
}

@article{fu2024symmetry,
  title={Symmetry classification of 2D materials: layer groups versus space groups},
  author={Fu, Jingheng and Kuisma, Mikael and Larsen, Ask Hjorth and Shinohara, Kohei and Togo, Atsushi and Thygesen, Kristian S},
  journal={2D Materials},
  volume={11},
  number={3},
  pages={035009},
  year={2024},
  publisher={IOP Publishing}
}

@article{Pakdel2023EmergentStacking,
  title={High-throughput computational stacking reveals emergent properties in natural van der Waals bilayers},
  author={Pakdel, Sahar and Rasmussen, Asbj{\o}rn and Taghizadeh, Alireza and Kruse, Mads and Olsen, Thomas and Thygesen, Kristian S},
  journal={Nature Communications},
  volume={15},
  number={1},
  pages={932},
  year={2024},
  publisher={Nature Publishing Group UK London}
}

\end{document}